\documentstyle[12pt,amsfonts,amssymb,amscd]{report}
\begin{document}
\begin{titlepage}
\pagestyle{empty}
\title{The Godbillon-Vey class, invariants of manifolds and linearised M-Theory}
\author{Ioannis P. \ ZOIS\thanks{izois\,@\,maths.ox.ac.uk and
izois\,@\,cc.uoa.gr; Research supported by the EU, contract no HPMF-CT-1999-00088}\\
\\
Mathematical Institute, 24-29 St. Giles', Oxford OX1 3LB\\}
\date{}
\maketitle
\begin{abstract}

We apply the Godbillon-Vey class to compute the transition amplitudes between some non-commutative solitons in M-Theory; our context is that of Connes-Douglas-Schwarz where they considered compactifications of matrix models on noncommutative tori. Two important consequences follow: we describe a new normalisation for the Abelian Chern-Simons theory using symplectic 4-manifolds as providing cobordisms for tight contact 3-manifolds and we construct a new(?) invariant for 3-manifolds. Moreover we modify the topological Lagrangian density suggested for M-Theory in a previous article to a \textsl{quadratic} one using the fact that the \emph{functor of immersions is a linearisation (or ``the differential'') of the functor of embeddings}.\\


 
\end{abstract}
\end{titlepage}

\section{Some remarks on K-Theories}

Let $M$ be a closed smooth $n$-manifold. A codimension $q$ \emph{foliation} where $0<q<n$, is a \textsl{particular example} of a \emph{Haefliger} or $\Gamma_q$-structure on $M$. A codim-$q$ foliation is defined by specifying a codim-$q$ \emph{integrable subbundle} $F$ of the tangent bundle $TM$ of $M$. A local definition is provided by specifying a nowhere vanishing decomposable $q$-form $\Omega$ on $M$. The integrability condition is expressed by the relation
$$\Omega \wedge d\Omega =0$$
or equivalently by the relation
$$d\Omega =\theta \wedge \Omega$$
for \emph{some} 1-form $\theta $ whose role will become clear later.\\

We would like to ask the following question:\\

Given a codim-q foliation on $M$ (namely an $F$ or an $\Omega $ as described above), how many K-Theories can be involved in our discussion?\\

The answer is \emph{six (6)}. We shall try to clarify them and show how they are related.\\

1. By definition the codim-q integrable subbundle $F$ of $TM$ defines a \textsl{class} in $K^0(M)$, where $K^0(M)$ is Atiyah's ``ordinary'' K-Theory, namely the abelian group consisting of stable isomorphism classes of vector bundles over $M$ (see for example \cite{Atiyah}). This $K-functor$ is represented on the ordinary de Rham cohomology groups of $M$ (just via the Chern-Weil theory) and one can use pairings with ordinary homology (simply integration of forms over suitable submanifolds of $M$) and get invariants. There are also some ``twisted'' versions of this, see \cite{Kar}. Their relation to D-Brane physics was discussed in \cite{WT} (this includes the torsion case; for a discussion for the non-torsion case see \cite{Bou} where they consider the noncommutative $C^*$-algebra $C(M)\otimes K$ where $C(M)$ is just functions on the manifold $M$ and $K$ is the elementary $C^*$-algebra of compact operators; this noncommutative $C^*$-algebra is in fact \emph{Morita equivalent} to the commutative algebra $C(M)$ as was clearly exhibited in \cite{Zois}). The key idea is that Grothendieck's stability relation corresponds to creation or annihilation of D-brane-antibrane pairs. For a compact Lie group $G$ one can have \textsl{G-equivariant} versions both of ordinary K-Theory and of twisted K-Theory (this is mainly due to Atiyah and Segal from late '60's). The later is closely related to the Verlinde algebra of $G$. In \cite{Tele} the authors claimed that they have ``almost'' proved that they actually coincide ``on the nose''. Moreover Gepner showed that the Verlinde algebra of $U(k)$ coincides with the quantum cohomology of the Grassmannian $G(k,N)$ at level $N-k$ (see for instance the review paper \cite{WittenY} and references therein).\\

\emph{2. A new K-Theory}\\
To the best of our knowledge,  \textsl{this is the first time that \emph{this} K-Theory appears in the literature} although the definitions follow essentially from the work of \cite{Haefliger}, \cite{Segal}, \cite{Bott} and \cite{Quillen}. The basic definitions are given in the last section (Appendix). In a way which is analogous but more complicated than the case of bundles, the set of all Haefliger structures of $M$ modulo some equivalence relations gives rise to another K-Theory, which we shall denote $K_{\Gamma}(M)$. We shall describe the strategy of this construction which is due to Quillen because we shall use it again later; roughly it goes as follows: given any category $C$ with a composition law, one can construct a representable contravariant functor $K_C$ from topological spaces to abelian groups which also preserves addition. The K-groups of the category $C$ are defined as the values of the functor $K_C$ on spheres, thus they are the homotopy groups of the space which represents $K_C$. This K-Theory, in striking contrast to 1. above does not have Bott periodicity; it follows the rules of Quillen's ``Higher Algebraic K-Theory''.\\

We apply this general construction in our case as follows: for a given manifold $M$, the space of all Haefliger structures modulo homotopy invariance and an equivalence relation (which is analogous to ``gauge transformations'' or ``change of trivialisation'' for bundles) denoted $\Gamma (M)$, forms a topological category. It is the analogue of the space of all (say complex) vector bundles over $M$ modulo gauge equivalence $Vect (M)$ of 1. above (see \cite{Atiyah}). The difference is that whereas $Vect (M)$ is an \textsl{abelian semigroup} and to get an abelian group one has to divide simply by the diagonal action (or equivalently impose Grothendieck's stability relation), $\Gamma (M)$ does not have this property and it is only a topological category. That category plays the role of the category $C$ and then we just apply the Quillen-Segal construction and get the abelian group $K_{\Gamma (M)}(M)$. Hence with a little more effort one can indeed ``concoct'' an abelian group finally in this case also.

Since foliations constitute a particular example of Haefliger structures, a foliation defines a \textsl{class} in that K-Theory too. We restrict ourselves to the 0th K-group of this K-Theory. Now this $K_{\Gamma}$-functor can be represented on the ordinary de Rham cohomology groups of our manifold $M$ but the appropriate cohomology (the analogue of Chern-Weil theory) is the \emph{Gelfand-Fuchs} cohomology (see \cite{Zois} and references therein). One can then use pairings with ordinary homology (just integration of differential forms) and get invariants. \emph{This is the framework in which our application falls in this article.}.\\

3. Given a foliation $F$ of our manifold $M$, one can construct the \emph{graph} or \emph{holonomy groupoid} denoted $G(F)$ which is due to Wilnkenkempern (see \cite{Wilnkenkempern}).
 This can also be seen as a topological category (it is also a manifold of dimension $dimM+dimF$, not necessarily Hausdorff though). One can also then apply the Quillen-Segal construction with $G$ playing the role of the category $C$ in 2. above. Now the foliation defines \emph{a whole K-Theory} and not just a class! We shall denote this K-Theory $K_G(M)$ and that's what's denoted in Conne's book
    $K^{*}_{top}(G)$.\\

4. In Conne's book there is an \textsl{alternative description} of the above, using the notion of \emph{``geometric cycles''}, namely K-oriented maps from compact manifolds to the space of leaves of the foliation; in more concrete terms a geometric cocycle is a triple $(W,y,g)$ where $W$ is a compact manifold, $y$ is an element of $K^{0}(W)$ (namely a bundle over $W$) and $g$ is a smooth map from $W$ to the space of leaves of the foliation (this is the key ingredient and roughly one can think of $G$ as the space of leaves of the foliation). For the precise definition see \cite{Connes}. One defines an equivalence relation between these geometric cycles (see again \cite{Connes} p126). The reason why  we mention this here is because it clarifies one important feature perhaps useful in physics (what we have in mind is of  course what's called ``F-Theory'' or perhaps any even higher dimensional theories that may appear in the future; in F-Theory the spacetime manifold has dimension 12): \emph{the dimensionality of the manifold $W$ representing the geometric cycle is not fixed but it can vary!} (this simply means that different foliations on different manifolds may have the same graph).\\

5. The holonomy groupoid $G$ of our foliation can be made into a $C^*$-algebra 
denoted $C^*(G)$,
using the vector space of \textsl{half-densities} over the manifold $G$ and completing it in a suitable way. This $C^*$-algebra then also has a K-Theory which is defined as stable isomorphism classes of finitely generated projective modules of the $C^*$-algebra and this has Bott periodicity (namely there are only two of these K-Groups). We shall denote this K-Theory $K_{*}(C^*(G))$. This $K$-functor can be represented in the \emph{cyclic cohomology} of the corresponding $C^*$-algebra and to get invariants one must use the pairing introduced by Connes (see \cite{Connes}) which involves the \emph{cup product \#} in cyclic cohomology. This is the real noncommutative framework. It is more complicated and it is what was used in \cite{Zois} to produce \emph{new} invariants.\\
 
 To be more specific, in this case one 
has two groups, namely $K_0$ which by definition is the
"Grothendiek group" $Gr$ of
$$K_0(C^*(G)):=Gr\pi _{0}P(C^*(G))$$ 
where $\pi _{0}$ as usual denotes the connected component and
$$P(C^*(G)):=\lim_{n \rightarrow\infty}Proj M_n(C^*(G))$$
$M_n$ denotes $n\times n$ matrices with entries from $C^*(G)$ and $Proj$ means that they satisfy the projectivity relation $e^{2}=e$ where $e$ an element of the algebra (these elements are also called ``idempotents'').
 An equivalent description is the following:

$$K_0(C^*(G))=\pi _{1}GL(C^*(G))$$
where $GL$ denotes the inductive limit for $n\rightarrow \infty $ of $GL(n;C^*(G))$. 

 Moreover, 
$$K_{1}(C^*(G)):=\pi _{0}GL(C^*(G))$$

For more details one can see \cite{Wegge}.\\ 

6. Following  an original idea due to Atiyah or the K-Homology according to Baum-Douglas, (see \cite{Baum-Douglas}),
 one can construct an \emph{alternative description} of K-classes of 4. above by using triples $(H, \pi ,T)$ of \emph{even} or \emph{odd} \emph{Fredholm modules}, where $\pi $ is an involutive representation of the $C^*$-algebra of the foliation $C^*(G)$ to a Hilbert space $H$ and $T$ is an operator satisfying the following requirements: it is self adjoint of modulus 1 and $[T,\pi (a)]$ is a compact operator for all elements $a$ of the $C^*$-algebra (see \cite{Connes} for full details, chapter 4). What we have just described is actually an \textsl{odd} Fredholm module; to define an \textsl{even} Fredholm module one needs an odd Fredholm module plus a $Z/2$-grading $\gamma $ of the Hilbert space $H$ satisfying the following requirements:\\
$\gamma =\gamma ^*$\\
$\gamma ^{2}=1$\\
$\gamma \pi (a)=\pi (a)\gamma $ for all $a$ in the $C^*$-algebra $C^*(G)$\\
$\gamma T=-T\gamma$\\
 These cycles are also represented in the \emph{cyclic cohomology} groups of the corresponding $C^*$-algebra of our foliation $C^*(G)$ (or the \emph{entire} cyclic cohomology if one in addition uses the property of \emph{$\theta $-summability} of the Fredholm modules, for a detailed exposition see \cite{Connes}).\\

As we already mentioned, 3. and 4. are actually two different descriptions of the same theory; the same is true for 5. and 6. following from Baum-Douglas' $K$-Homology. That the 3. (or its equivalent 4.) and 5. are \emph{isomorphic} (the 0th and 1st groups for 3.) is the famous \emph
{Baum-Connes conjecture} which includes the \textsl{Novikov conjecture} for higher signatures of manifolds as a special case.
The 1. is well understood and 2. is what we shall use in this piece of work to give an application in M-Theory based on the Connes-Douglas-Schwarz compactifications of Matrix theory to noncommutative tori (\cite{Connes-Douglas-Schwarz}). 

So one can say that essentially there are two interesting K-Theories involved, namely 2. and 5. (assuming the Baum-Connes conjecture is true). One could describe them simultaneously using a \emph{bifunctor} but this is not needed in our work here. Good  candidates are Kasparov's $KK$-bifunctor (see \cite{Kasp}) or even better Connes' $E$-bifunctor which has a better behaviour than $KK$ (for example $E$ is half exact whereas $KK$ is not, see \cite{Connes}).

\section{The Godbillon-Vey class}

Given a codim-q foliation $F$ on our smooth closed $n$-manifold $M$, one can define the \emph{normal bundle} $Q$ of the foliation as $Q=TM/F$. Clearly $Q$ is a rank-q vector bundle over $M$. \textsl{Since $F$ is integrable, the Pontryagin classes of $Q$ of degree grater than $2q$ vanish} (for the proof see \cite{Bott}).\\

Let $C^{\infty }(F)$ denote the set of sections of the vector bundle $F$. The Frobenious theorem states that for two vector fields $X$ and $Y$ in $C^{\infty } (F)$, their commutator $[X,Y]\in C^{\infty } (F)$ (this follows from the integrability of $F$). In an obvious notation then, if $Z\in C^{\infty } (Q)$, then $Z=\pi (Z')$ for some $Z'\in C^{\infty } (TM)$, where $\pi :TM\rightarrow Q$ is the canonical projection. $Z'$ is well defined modulo elements of $C^{\infty } (F)$. Thus for $X\in C^{\infty } (F)$ and $Z\in C^{\infty } (Q)$, one can define
$$\nabla _{X}(Z):=\pi [X,Z']$$
This is clearly an ${\bf R}$-bilinear map
$$\nabla :C^{\infty } (F)\times C^{\infty } (Q)\rightarrow C^{\infty } (Q)$$
and satisfies
$$\nabla _{X}(fZ)=X(f)Z+f\nabla _{X}(X)$$
and
$$\nabla _{fX}(Z)=f\nabla _{X}(Z)$$
as is easily verified. This ``almost'' satisfies the definition of a \emph{connection} on $Q$ except that the variable $X$ is restricted to range over $C^{\infty } (F)$ instead of over all of $C^{\infty } (TM)$. In order to \textsl{complete} it to a connection, one can either use a Riemannian metric on $M$ or another full connection on $Q$. Under the assumption that $F$ is integrable, one can prove that $Q$ has a connection of this kind. (For more details see \cite{Bott} or \cite{Kamber}). This is called a \emph{basic} connection (or \emph{Bott} connection) on $Q$.\\

We now restrict ourselves for the moment to the case where $F$ is of codim-1. Hence locally $F$ is defined by a nowhere vanishing 1-form $\Omega $. One can then prove that
$$d\Omega =\theta \wedge \Omega $$
if and only if $\theta $ is the connection 1-form of a basic connection on $Q$. (For the proof see \cite{Bott}).\\

By definition, a basic connection has an important property: its curvature denoted $K_{\theta }$ which by definition is as usual 
$$K_{\theta }:=d\theta  +\theta \wedge \theta =d\theta $$
namely the nonlinear term $\theta \wedge \theta $ vanishes! (for the proof see \cite{Bott}). This makes basic connections \emph{look like being abelian}. (This is true for a codim-q foliation also).\\

{\bf Aside:}
 We would like to make another remark here: in the complex case there is a correspondence between \emph{flat} connections (which correspond to ``local coefficient systems'') and semistable \emph{Higgs bundles} with vanishing 1st and 2nd Chern classes (the notion of a Higgs bundle was introduced by Hitchin and is a bundle with a Higgs connection which by definition behaves like a basic connection). For the proof see \cite{Simp}.\\

Then the \emph{Godbillon-Vey class} is exactly the class defined by the 3-form $\theta \wedge d\theta $. One can prove that it is closed, independent of the choice of the connection 1-form $\theta $ and that it characterises codim-1 foliations up to homotopy (for the proofs of these statements see \cite{Bott}).\\

One can in fact generalise this construction for any foliation of codim-q; then the corresponding class will be a $(2q+1)$-form 
$$\theta \wedge (d\theta )^{q}$$ 
where the power means wedge product. A codim-q foliation $F$ of a smooth closed $n$-manifold $M$ is defined locally by a nowhere vanishing decomposable q-form $\Omega $ and integrability means that $d\Omega  =\theta \wedge \Omega $ if and only if $\theta $ is the connection 1-form of a \emph{basic} connection on the normal bundle $Q=TM/F$ of the foliation (as we had stated in the start of the previous section). This class belongs to  the Gelfand-Fuchs cohomology as was discussed in a more detailed way in \cite{Zois}. In fact it is the only known non-trivial class of the Gelfand-Fuchs cohomology which characterises foliations up to homotopy. 

We would like to use this class in the case of noncommutative compactifications of matrix models in M-Theory as described in \cite{Connes-Douglas-Schwarz} (see also \cite{SW}, \cite{Nek1}, \cite{Nek2} and \cite{Kap}).

\section{Noncommutative vacua}

The idea that noncommutative geometry might be of relevance in M-Theory was pointed out for the first time in \cite{Connes-Douglas-Schwarz}. In fact these authors constructed explicitly compactifications of the matrix models into noncommutative tori. The framework was that of operator algebras but as was pointed out in subsection 4.1 of that paper an equivalent description exists involving foliations. In fact the meaning of a ``noncommutative torus'' is that of a torus with a foliation whose corresponding $C^*$-algebra is noncommutative.

The construction of the particular solutions of certain equations (which correspond to compactifications of the matrix model) in that paper can be described as a two step procedure:\\

1. First one has to specify a noncommutative algebra denoted $T_C$.\\

2. Then the second step is to construct explicit modules of this algebra $T_C$ which can be thought of as ``bundles with connections'' over that ``quantum space''.\\ 

The physical interpretation given to the above process and its relation to M-Theory was the following: these solutions (noncommutative compactifications) were shown to correspond to supersymmetric vacua (namely elements of the moduli space) of D=11 supergravity, assuming the 11-dim manifold to be of the form $T^{d}\times M^{1,10-d}$ where the notation means that $T^d$ is a $d$-dim torus and $M^{1,10-d}$ is an $(11-d)$-manifold with 1 time coordinate and $(10-d)$ spatial coordinates of Minkowski type.\\

Now our application starts with the following question:\\ 

\emph{Can these vacua interpolate?}\\

What we have in mind of course is the case of instantons, where quantum mechanically an instanton is exactly an interpolation between classical \emph{gauge inequivalent} solutions of the Yang-Mills equations. Essentially they correspond to topologically distinct G-bundles over spacetime, assumed to be ${\bf R}^4$ compactified to $S^4$.\\

Let us start with fixing the dimension $d$ for the torus in our 11-manifold with its Cartesian product decomposition as describd above. Then it is clear we believe that the K-Theory which is of relevance for the first step above is the second in our list which, remember, contains the different Haefliger structures (and hence includes foliations) of our torus, since the choice of a noncommutative algebra $T_C$ means that we have chosen a foliation of our torus.\\

Having done that, the second step is to construct modules or equivalently bundles with connections over that quantum space; the K-Theory relevant to this step is clearly no 5. in our list in the first section which contains stable isomorphism classes of finitely generated projective modules of the noncommutative algebra $T_C$.\\

For a fixed dimension $d$ of the torus, one might say that in fact there are \textsl{two} \emph{``levels''} of interpolation: namely either between different $T_C$'s (i.e. between different--up to homotopy--foliations of our torus $T^{d}$, hence essentially ``counting'' $K_{\Gamma }(T^{d})$ classes) or within the same $T_C$ between different--up to Morita Equivalence--modules of $T_C$, hence essentially ``counting'' $K_{0}(T_{C})$ classes. Let us recall that there is actually an analogue of this in the commutative case when studying \emph{ALE instantons}, namely instantons where the boundary has non-trivial fundamental group.\\

The second is more elaborate and in fact we do not want to comment extensively in this article on how one could compute the transition amplitudes in this case (see however the comment in the last paragraph of this section below which describes the key idea). But for the first, the idea is that one can use the Godbillon-Vey class as a Lagrangian density integrated over the fundamental class of our manifold to get an action functional and try to calculate the relevant path integral. It is reasonable to expect that this will give the contribution from interpolation between non-homotopic foliations (hence according to the Connes-Douglas-Schwarz physical interpretation that would amount to interpolation between homotopically distinct noncommutative vacua of D=11 SUGRA). But one has to make special choices in order to be able to calculate the path integral: the dimension of our torus $d$ has to be 3 and we can calculate only the transition amplitude between codim-1 foliations of the $T^3$ (namely, for \emph{some} of the possible noncommutative algebras $T_C$'s, those coming from codim-1 foliations of the $T^3$). The reason for that is because for $codim > 1$ the Godbillon-Vey class is \textsl{not quadratic} in $\theta $ and one would have to use \textsl{some} approximation. Since the codim has to be 1, then the Godbillon-Vey class is a 3-form and hence in order to be used as a Lagrangian density one must have a 3-manifold, hence $d$ has to be 3. However our argument can be applied to any closed smooth oriented connected 3-manifold, namely not just to toroidal compactifications. There is a hope that one might be able to use stationary phase approximation for the 5-dim case since in this situation the Godbillon-Vey class is $\theta\wedge (d\theta )^{2}$ which is certainly not quadratic but perhaps managable.\\

Before starting this calculation, we would like to make a comment about the Godbillon-Vey class: every fibre bundle is a foliation (of the total space of the bundle; the fibres are the leaves); consider the first Hopf fibration $S^{2}\hookrightarrow S^{3}$ seen as a foliation of $S^{3}$ with leaves being $S^{2}$. Then the Godbillon-Vey class for this codim-1 foliation is the topological term added to the usual principal chiral $\sigma $ model action to get the \emph{skyrmions} in \cite{Zee}. The topological charge of skyrmions (namely the integral of the Godbillon-Vey class over $S^{3}$) is ``more or less'' the Hopf index. (More or less because the principal chiral $\sigma $ model has as target space the Lie group $SO(3)$ which is topologically half of $S^{3}$). For the case of a general compact connected Lie group $G$ as being the target space (instead of just $SO(3)$), one has the so called ``Wess-Zumino'' invariant (or term). Generalising the source space of the $\sigma $ model to a Riemann surface with genus grater than 0, replacing $S^2$ that is, as well as with its relation to Chern-Simons invariant, see \cite{Freed}. So the Godbillon-Vey class should be thought of as a noncommutative generalisation of the above really defined for \emph{any} codim-1 foliation. In the special examples mentioned above it reduces to well-known invariants (namely the Hopf invariant for the 1st Hopf fibration and to the Wess-Zumino invariant for $\sigma $ models with source any Riemannian surface and target a Lie group).\\

We would like to finish this section with the key observation which in principle enables one to compute transition amplitudes between different modules over a fixed quantum space (namely the second level of interpolation according to our terminology above); in the path integral computations essentially the only Lagrangians whose partition function can be computed are the quadratic ones and what actually comes out as the result is the determinant of an operator (this will be quite explicit in the next section). The path integral expression itself does not make much sense \emph{a fortiori} when one considers a quantum topological space and wants to ``integrate'' a Lagrangian density over it to get the action (namely equation $(*)$ in the next section). Equation $(**)$ though as we shall see in the next section in our case involves almost exclusively the Laplace operator of the manifold considered. The point then is that there is an analogue of the Laplace operator (which in fact equals $1-\kappa $ where $\kappa $ is the \emph{Karoubi} operator) which gives the analogue of \emph{Hodge theory} for the cyclic cohomology of any associative algebra (clearly in our case the associative algebra will be the algebra $T_C$ of the foliation). Then as regularised determinant one should use the $exp$ of the \emph{Dixmier} trace. So equation $(**)$ in the next section makes \emph{perfect sense} even over quantum spaces! The analogue of Hodge theory for the cyclic cohomology of any associative algebra is due to Cuntz and Quillen (see \cite{Quillen}).

\section{The computation}

We shall recall some facts first about degenerate quadratic functionals. Our references in this section are \cite{S1}, \cite{S2} and \cite{S}.

A non-negative self-adjoint operator $B$ in a Hilbert space is called \emph{regular} if for $t\rightarrow +0$ one has
$$Tr(exp(-Bt)-\Pi (B))=\sum a_k(B)t^{-k} + O(t^{\epsilon })$$
where $\epsilon >0$ and $k$ runs over a \emph{finite} set of non-negative integers. (All operators will be assumed as being elliptic). In fact since we shall be considering operators acting on sections of some vector bundle over a closed smooth Riemannian $n$-manifold $M$, then $k$ will take integer values from 0 to $n$, (see for example \cite{Roe}) and in this case the coefficients $a_k$ can be calculated by local expressions using the Seeley formulae, see \cite{S}. The symbol $\Pi $ denotes the projector on the kernel of the operator considered and $Tr$ denotes the usual trace.\\

We say that the operators $A$ and $B$ acting in a Hilbert space $H$ form a \emph{regular pair} if $B$ is a non-negative self-adjoint operator and for $t\rightarrow +0$ one has
$$TrA(exp(-Bt)-\Pi (B))=\sum a_{k}(A|B)t^{-k} + O(t^{\epsilon })$$
and for $t\rightarrow \infty $ one has
$$TrA(exp(-Bt)-\Pi (B))=O(t^{-N})$$
where $\epsilon >0$ and $N$ is an arbitrary integer. If $A=1$ then the pair $(A,B)$ is regular if and only if $B$ is regular.

The \emph{zeta } function $\zeta (s|B)$ of the operator $B$ for large $Re(s)$ can be defined by the formula:
$$\zeta (s|B):=\sum \lambda _{j}^{-s}=\frac{1}{\Gamma (s)}\int _{0}^{\infty }Tr (exp (-Bt)-\Pi (B))t^{s-1}dt$$
where $\lambda _{j}$ are the non-zero eigenvalues of $B$. Then we define the \emph{regularised determinant} $D(B)$ of the regular operator $B$ to be
$$log D(B):=-\frac {d}{ds}\zeta (s|B)|_{s=0}$$
This definition is correct because the zeta function is analytic at the point $s=0$. In an analogous manner one can define \emph{families} of regular operators (see for example \cite{S1}).\\

Let $L$ be a \emph{quadratic} functional on a Hilbert space $H$, namely
$$L(f)=<Sf,f>$$
where $S$ is a self-adjoint operator acting on $H$, $<,>$ denotes inner product in $H$ and $f\in H$. The functional $L$ is called \emph{non-degenerate} if $Sf=0\Leftrightarrow f=0$. If $S^2$ is regular, then we define the \emph{partition function} $Z$ of $L$ as follows:
$$Z(L):=D(S)^{-1/2} = D(S^2)^{-1/4}$$
where $D$ denotes the regularised determinant.\\

Now suppose that our functional is \emph{degenerate}, namely there exists a linear operator $T$ on the Hilbert space $H$ such that:
$$L(f+Th)=L(f)$$
where $f,h \in H$. One can check that this requirement is satisfied if and only if $ST=0$. Assuming that there exists an adjoint operator $T^*$ of $T$ and that both operators $S^2$ and $T^{*}T$ are regular,  we can define the \emph{partition function} $Z$ of a degenerate now functional $L$ by the relation:
$$Z(L):=D(S)^{-1/2}D(T)$$
The origin of this definition as we think is quite clear, is the \emph{``Fadeev-Popov'' trick} in quantum gauge theory (or \emph{BRST-formalism} as it is known in physics) and we do not consider the delicate question of Gribov ambiguities.\\

It is not hard to rewrite the partition function in a more convenient form (see \cite{S1}):
$$ Z(L)=D(\Box _0)^{-1/4}D(\Box _1)^{3/4}$$
where
$$\Box _{0}:=S^{2} + TT^{*}$$
and
$$\Box _{1}:=T^{*}T$$

Now let us assume that we have a smooth closed $n$-manifold $M$ equipped with a Riemannian metric and we consider the \emph{elliptic deRham complex} of differential forms over $M$ in the usual way, as described for example in \cite{A}. The Riemannian metric gives rise to inner product among differential forms, hence one can get a (pre)Hilbert space.\\

Then we would like to compute the partition function

$$(*) Z = \int D\theta  exp (iY \int _{T^3}\theta \wedge d\theta )$$

namely the partition function of the Godbillon-Vey class integrated over the 3-torus $T^3$ and $Y$ is a normalisation constant (in fact we choose $Y$ to be $k/8\pi $ so as to coincide with abelian Chern-Simons theory). Clearly, from our discussion above, the action is quadratic, the operator $S$ is just the deRham differential $d$. Our action functional $L = \int _{T^3} \theta \wedge d\theta $ is actually degenerate; one can very easily check that 
$$L(\theta +dc)=L(\theta )$$ 
Hence the operator $T$ in our general formalism described above is again the deRham differential $d$. Thus applying our definition for the partition function of degenerate quadratic functionals, the answer is (ignoring constants):
$$(**) Z_{T^3}=D(\triangle _{1})^{-1/4}D(\triangle _{0})^{3/4}$$
where $\triangle _{i}=d^{*}d+dd^{*}$ is the Laplace operator acting on the $i$-forms, $i=0,1$.\\

Let us forget the normalisation of the partition function for the moment. It seems as if the result depends actually on the  Riemannian metric of our 3-torus. However one can prove that this is not so, in fact we actually have a \emph{topological} quantum field theory (or a \emph{generally covariant} quantum field theory). To prove this we follow the general strategy described in \cite{S1}.\\

We introduce a continuous \emph{family} of Riemannian metrics $g_{u}$ on $T^3$ parametrised by the parameter $u$, where $u\in [0,1]$. (We assume that all families are smooth). Hence we actually get a \emph{family} of inner products $<,>_{u}$ among differential forms parametrised by the same parameter $u$, where again $u\in [0,1]$. The variation of the corresponding inner products by infinitesimal variation of the metrics can be described by means of a family of operators $B(u)$ defined as follows:
$$\frac{d}{du}<f,g>:=<B(u)f,g>=<f,B(u)g>$$
for $f,g\in H$ (in our case $f,g$ are actually differential forms but we keep the more general notation assuming that they actually belong to a Hilbert space $H$; such a space can be defined using the deRham complex of differential forms over a closed Riemannian manifold, see for example \cite{A}).

That in turn gives rise to a family of self-adjoint operators $S(u)$ and operators $T(u)$ which express the degeneracy (or the gauge freedom) of our action functional. Making the same assumptions as above concerning the regularity of these operators, we finally get a family of operators $\Box _{0}(u)$ and $\Box _{1}(u)$ defined in a similar fashion. Eventually we get a family $Z(u)$ of partition functions. Then one has the the following:\\

{\bf Proposition 1:}

$$\frac{d}{du}log Z(u)=\frac{1}{2}a_{0}(B(u)|\Box _{0}(u))-\frac{1}{2}a_{0}(B(u)|\Box _{1}(u)$$
where of course the coefficients $a_0$ are the ones appearing in the definition of a regular operator suitably generalised for pairs of regular operators in this case. These coefficients as we stated before, can be computed by local expressions using the Seeley formulae. For the proof of the proposition we refer to \cite{S1} (but in fact it follows from statements proved in \cite{S2}). It relies on direct but somehow tedious calculations.\\

But now we recall an important application of the Atiyah-Singer index theorem (see \cite{A}):\\

The index of an elliptic differential operator on an \emph{odd} dimensional smooth closed manifold is zero.\\

From the heat kernel proof of the Atiyah-Singer Index theorem we also know, using the Seeley formula once more, that the index of an elliptic differential operator say $A$ is actually equal to (let $A^*$ denote the adjoint of $A$):
$$ind(A)=a_{0}(AA^{*})-a_{0}(A^{*}A)$$
where the coefficient $a_{0}$ is again the one appearing in the definition of a regular operator. (See for example \cite{At} or \cite{Roe}). Since we are on a 3-manifold the index has to vanish and hence the derivative of our partition function which by the proposition was proved to be equall to the index of some operator on a 3-manifold vanishes too; hence our partition function is \emph{metric independent}.\\

We shall make various remarks now:\\ 

The first and most important one is about the \textsl{domain} of the path integral. In our discussion above we assumed integration over \emph{all} 1-forms $\theta $. Yet from our discussion about foliations, it seems more appropriate, in order to compute the contibution from interpolation between non-homotopic codim-1 foliations of the 3-torus, to consider only \emph{basic} connection 1-forms $\theta $. The problem we have of course in this case is that we do not know what this space is. If we were on an even dimensional manifold which could be equipped with a complex structure we could apply Simpson's result which establishes a  1:1 correspondence between basic (=Higgs) connections and ``flat'' connections (the spaces of flat connections have in general been studied more than basic connections). \textsl{The crucial observation though is} that if we restrict ourselves to a particular case of codim-1 foliations called \textsl{``taut''} (for the definition see the next section) and we want to compute the contribution to the path integral from interpolation between non-homotopic taut codim-1 foliations, we encounter an interesting result due to Thurston and Kronheimer-Mrowka, which states that for a closed oriented connected 3-manifold (provided a restriction holds), there is only a \emph{finite number} of those. In this case the path integral should be replaced by a \emph{finite sum} and that's certainly (in principle) calculable. \emph{This fact may be used to define new invariants for 3-manifolds}, perhaps related to invariants which are already known. More on this in the next section where we shall mention all the relevant details.\\

It is we think an interesting feature that at the quantum level this theory (describing codim-1 foliations on a 3-manifold) coincides with Abelian Chern-Simons theory \emph{if} we integrate over all connection 1-forms. The fact that we talk about foliations, hence basic connection 1-forms, seems to correspond to being ``on shell'' for the non-abelian Chern-Simons theory on a (2+1) 3-manifold, bearing in mind as a rough analogy the results proved in \cite{Simp}. Let us explain this point more: Recall that for non-abelian Chern-Simons theory on a 3-manifold, being on-shell means flat connections; the (2+1) decomposition of our 3-manifold needed for relating Lagrangian and Hamiltonian formalism essentially reduces the problem from our original 3-manifold to a 2-manifold whose space of flat connections has been extensively studied; this is the reason why the geometric quantisation scheme applies and in fact was used in \cite{WN} to solve non-abelian Chern-Simons theory; a 2-manifold can obtain a complex structure but then Simpson's result establishes a 1:1 correspondence between Higgs (=basic) connections and flat connections--in fact local coefficient systems (we have ignored some of the details of Simpson's results, namely that Higgs bundles have to be semistable with vanishing 1st and 2nd Chern classes). In the last section where we shall discuss M-Theory we shall see a striking similarity with Chern-Simons theory in dim 3.\\

A full treatment of both Abelian and non-Abelian Chern-Simons theory with its relation to knots and links was given in \cite{Gu} (see also \cite{Turaev}). In general there are \emph{two} approaches: \textsl{the observables expectation values'} approach and the \textsl{topological quantum field theory} approach.\\

We start using the first and in particular we would like to comment on the importance of ``framings'' of knots (Wilson-lines), especially in the quantum level. The introduction of framings is not motivated by the need to eliminate divergencies but by the necessity of preserving general covariance.\\

We shall use the physics terminology now (our previous discussion based on the concise terminology of \cite{S1} essentially captures BRST-formalism): in order to define the quantum theory of Abelian Chern-Simons theory in 3 dimensions, one can start by applying the well-known BRST formalism; in the Landau gauge the total action will be:
$$\tilde{L}=\frac{k}{8\pi }\int d^{3}x (\epsilon ^{\mu \nu \rho }\theta _{\mu }\partial _{\nu }\theta _{\rho }-B\partial ^{\mu }\theta _{\mu }+\bar{c}\partial ^{\mu }\partial _{\mu }c)$$
where $B$ is a bosonic auxiliary field and $\bar{c},c$ are the ghost fields. Since the metric enters in the gauge fixing procedure only, for gauge invariant and metric independent observables general covariance is preserved. One can compute the propagators
$$<\theta _{\mu }(x)\theta _{\nu }(y)>=\frac{i}{k}\epsilon _{\mu \nu \rho }\frac{(x-y)^{\rho }}{|x-y|^3}$$
$$<\theta _{\mu }(x)B(y)>=-\frac{i(x-y)_{\mu }}{k|x-y|^3}$$
and
$$<c(x)\bar{c}(y)>=-\frac{i}{k|x-y|}$$ 

 Since we can calculate the partition function one can go on and try to calculate some vacuum expectation values
$$<e^{i\int J\theta }>=Z^{-1}\int D\theta exp (i\int _{M^3} \theta \wedge d\theta )e^{i\int J\theta }$$
where $J$ is some source term. An important class of such terms which would have to be gauge invariant quantities are the so called Wilson lines $W(C)$, where $C$ is some loop in our 3-manifold or more generally an oriented knot (these are supposed to be our observables). In the abelian theory, at the classical level, these Wilson lines are just line integrals but in the non-abelian case one has to introduce \emph{path order} and the situation is more complicated. One can represent them as an infinite sum of iterated integrals (see \cite{Z1}). 

At the quantum level though, even in the Abelian case one has to face the following complication which essentially forces one to use ``framed'' Wilson operators: in considering Wilson line operators one has to exponentiate the integral of the 1-form $\theta $. Consequently one has to analyse the case in which the product of two or more $\theta _{\mu }dx^{\mu }$ integrals performed on the \emph{same} loop occurs. One may try to do that naively by defining:

$$(\oint _{C}\theta _{\mu }dx^{\mu })^{2}:=\int _{0}^{1}ds \int _{0}^{1}dt \dot{x}^{\mu }(s)\theta _{\mu }(x(s))\dot{x}^{\nu }(t)\theta _{\nu }(x(t))$$
where $x(s)$ parametrises $C$ and dot means derivative with respect to the relevant variable. With this naive definition of $(\oint _{C}\theta _{\mu }dx^{\mu })^{2}$, the vacuum expectation values are finite but general covariance is not maintained. This is a common problem in quantum level and the origin of this is the following fact: at the classical level knowing $\theta (x)$ is sufficient for determining its powers $\{\theta ^{2}(x),...\}$ yet at the quantum level things are more subtle; the field operator $\theta (x)$ is well defined in the sense that its correlation functions are well defined but this is not enough for uniquely determining what the quantum operators $\{\theta ^{2}(x),...\}$ mean.

It turns out that the correct definition in order to maintain general covariance is to use framings for knots, namely for each knot $C$ parametrised by $\{x^{\mu }(s);s\in [0,1]\}$ we intoduce a \emph{framing} $C_f$ parametrised by
$y^{\mu }(s)=x^{\mu }(s)+\epsilon n^{\mu }(s)$ where $\epsilon >0$ and $|\vec{n}(s)|=1$, where $n^{\mu }$ is a vector field orthogonal to $C$. Then one defines:
$$(\oint _{C}\theta _{\mu }dx^{\mu })^{2}:=lim_{\epsilon \rightarrow 0}\oint _{C}\theta _{\mu }dx^{\mu } \oint _{C_f}\theta _{\nu }dy^{\nu }$$
with the convention that the $\epsilon \rightarrow 0$ limit has to be taken after all the Wick contractions and integrations have been performed. Of course at the classical level both definitions coincide. Hence now expectation values of powers of Wilson lines on the same loop are generally covariant.\\

The definition then of the composite Wilson line operator by means of the framing procedure is quite natural: the naive expression for the associated Wilson line operator would be
$$W(C)=exp(i\oint _{C}\theta _{\mu }dx^{\mu }):=\sum _{n}\frac{i^n}{n!}\oint _{C} \theta _{\mu _1}(x_1)dx_{1}^{\mu _1}...\oint _{C}\theta _{\mu _n}(x_n)dx_{n}^{\mu _n}$$
Yet as explained above this may cause problems in maintaining general covariance. To be more precise, in a generic product
$$\oint _{C} \theta _{\mu _1}(x_1)dx_{1}^{\mu _1}...\oint _{C}\theta _{\mu _n}(x_n)dx_{n}^{\mu _n}=\int _{0}^{1}ds_{1}...\int _{0}^{1}ds_{n}\dot{x}^{\mu _1}(s_1)\theta _{\mu _1}(x(s_1))...\dot{x}^{\mu _n}(s_n)\theta _{\mu _n}(x(s_n))$$
appearing in the previous equation, each term $\dot{x}^{\mu _i}(s_i)\theta _{\mu _i}(x(s_i))$ is replaced by
$$\dot{x}^{\mu _i}(s_i)\theta _{\mu _i}(x(s_i))\rightarrow [\dot{x}^{\mu _i}(s_i)\theta _{\mu _i}(x(s_i))]_{f}$$
where
$$[\dot{x}^{\mu _i}(s_i)\theta _{\mu _i}(x(s_i))]_{f}=(\dot{x}^{\mu _i}(s_i)+\epsilon _{i}\dot{n}^{\mu _i}(s_i))\theta _{\mu _i}(x(s_i)+\epsilon _{i}n(s_i))$$
where the vector field $n^{\mu }$ characterises the choice of framing as specified above and the values $\{\epsilon \}=\{\epsilon _{i}; \epsilon _{i}>0\}$ can be arbitrarily chosen provided $\epsilon _{i}\neq \epsilon _{j}$ for $i\neq j$.\\

Finally the composite Wilson line operator associated with a framed knot $C$ is defined as
$$W(C)_{f}:=lim_{\{\epsilon \}\rightarrow 0}\sum _{n}\frac{i^n}{n!}\int _{0}^{1}ds_{1}...\int _{0}^{1}ds_{n} \{[\dot{x}^{\mu _1}(s_1)\theta _{\mu _1}(x(s_1))]_{f}...[\dot{x}^{\mu _n}(s_n)\theta _{\mu _n}(x(s_n))]_{f}\}$$

In the non-abelian case one can define composite Wilson line operators associated with framed knots similarly.\\

Anyway we present for completeness the solution to the Abelian Chern-Simons theory; consider a generic oriented framed link $L$ with $m$ components $\{C_1,...,C_m\}$ in which the i-th component $C_i$ has framing $C_{if}$ (we omit the charges for simplicity). In each term of the expression above for the composite Wilson line operator the Wick contractions are defined using the propagators we had previously computed. As it is well-known the combinatorics of the resulting Feynman diagrams reproduce the expansion of an exponential term. The sum of all the contributions defined by the perturbative expansion gives the expectation value of the associated Wilson line operator $W(L)$:
$$<W(L)>=exp\{-i\frac{2\pi }{k}[\sum _{i}\chi (C_{i},C_{if})+2\sum _{i<j}\chi (C_{i},C_{j})]\}$$
where $\chi $ denotes the linking number. Hence we see that $<W(L)>$ represents a regular isotopy invariant for link diagrams.\\
 
In the framework of the second approach (the \emph{topological one}), one wants to see the Abelian Chern-Simons theory as a rule to assign invariants on the 3-manifold considered. In order to do that one should \emph{normalise} the partition function by computing it on a given fixed 3-manifold, say $S^3$, and then consider the ratio
$$R(M,S^3)=\frac{Z(M)}{Z(S^3)}=\frac{<0|0>|_{M}}{<0|0>|_{S^3}}$$
for any oriented closed connected 3-manifold $M$ where $Z$ is the partition function
$$Z(M)=\int _{M}D\theta exp(iL)$$
using the Abelian Chern-Simons action $L$. The partition function represents the vacuum to vacuum amplitude $<0|0>|_{M}$; this by itself cannot represent any meaningful invariant, unlike the expectation values of the observables
$$<W(L)>|_{M}=\frac{<0|W(L)|0>|_{M}}{<0|0>|_{M}}$$
which are well-defined because they are properly normalised.\\

These two approaches to 3-dim Chern-Simons theory are different: the observables expectation values approach described previously tells us how to compute the observables in a given 3-manifold only whereas for the topological quantum field theory approach one must have a way to obtain other 3-manifolds  from $S^3$ which is the fixed 3-manifold used for the normalisation of the partition function. To do that, there are two strategies: either move to dim 4 and consider \emph{cobordisms} or perform \emph{surgeries} on $S^3$ and get other 3-manifolds.\\ 

The first strategy has not been developed since there are many difficulties. Yet our original new idea along these lines here is that \emph{taut foliations give cobordisms between symplectic 4-manifolds}. In order to establish this \textsl{new} normalisation for the topological approach to 3-dim Abelian Chern-Simons theory (at least in the special case where the 3-manifold has a taut codim-1 foliation) we need to review some recent results from geometry in the next section where we shall mention ``symplectic fillings'' of 3-manifolds.\\

 As about the usual strategy of normalisation, namely using surgeries, we would like to mention simply that it leads (after inserting a non trivial phase factor to restore invariance under Kirby moves--this is not required in the Abelian case and it is the phase in \cite{WN} section 2 appearing from the Chern-Simons invariant of the \textsl{flat} connections which are the stationary points of the action) to the Reshetikhin-Turaev invariant for 3-manifolds (which is the rigorous version of Jones-Witten theory, see \cite{Res}). It is quite helpful that in fact these surgeries can be understood in terms of \emph{symmetry} transformations acting on the system. Moreover a surgery operator representing surgery actually exists so the result of surgery can be obtained by inserting that operator in the expectation values (hence one might say that the 2 approaches are related). We do not intend to repeat all the details here, they can be found in the textbooks which we refered to.\\ 

 The two theories (namely abelian Chern-Simons and Godbillon-Vey) are very closely related because considering codim-1 foliations on a 3-manifold gives a normal bundle which is of rank 1, hence one can think of it as an abelian adjoint bundle. For codim grater than 1 of course the two theories do not coincide but in this case the Godbillon-Vey class is not quadratic. But we think that there is a nice geometric interpretation (at least in one case which we will call \emph{transverse} instanton) describing these ``noncommutative solitons'', namely the interpolation between homotopically distinct codim-1 foliations of $T^3$ (or equivalently the interpolation between non Morita equivalent noncommutative $C^*$-algebras $T_C$  in \cite{Connes-Douglas-Schwarz}). We shall try to describe this in the following section. 

Let us also make another remark concerning our terminology: it is clear we think from the discussion above that we call a noncommutative instanton an interpolation between homotopically distinct codim-1 foliations of the 3-torus. It is unfortunate that in the literature the term noncommutative instanton is used to describe also deformations of ordinary instantons, namely instantons over ${\bf R}^4$ where the commutative algebra of functions on ${\bf R}^4$ is replaced by a deformation involving a star product (quantum algebra of the Moyal bracket). What we describe here is different.\\

The last remark in this section is the following: on a 3-manifold, say $M$, (closed, orientable, connected) we have 3 different 3-forms: 
$$A\wedge dA$$ 
namely the Abelian Chern-Simons 3-form, where $A$ is a connection 1-form on an Abelian bundle $E$ over $M$ (hence $A$ is a real valued 1-form); 
$$\Omega \wedge d\Omega $$ 
where $\Omega $ is a nowhere vanishing real 1-form which defines a codim-1 foliation $F$ on $M$. From the integrability of $F$ follows that $\Omega \wedge d\Omega =0$ or equivalently $d\Omega =\theta \wedge \Omega $ for $\theta $ a basic connection 1-form on the normal bundle $Q=TM/F$ of the codim-1 foliation; one then also has the Godbillon-Vey class of the codim-1 foliation $F$ which is 
$$\theta \wedge d\theta $$
In principle these 3-forms represent three different things, but they are related;
If the \emph{Abelian Chern-Simons} action is {on shell}, then $A$ is flat, hence since we are in the Abelian case this means that $A$ is closed, so the 3-form no1. above which represents the Chern-Simons invariant vanishes. Hence now the Abelian Chern-Simons action itself satisfies the integrability condition and $A$ (which is diferent from zero) can be used to define a codim-1 foliation on $M$. A \textsl{closed} 1-form as we shall see in the next section can define a codim-1 foliation which \textsl{cannot be linearly deformed into a contact structure}. Moreover in this case $\theta $ has to vanish, hence the Godbillon-Vey class is also zero. In general any $G$-bundle (where $G$ a compact connected Lie group) over a smooth closed $n$-manifold $M$ say gives a foliation of the total space of the bundle (just because the Lie algebra of \emph{vertical} vector fields, namely the vector fields tangent to the gauge orbits, closes). If in addition this $G$-bundle is \emph{flat} one has an extra foliation called \emph{horizontal}. This case was extensively studied in \cite{Zois}; the relation between the Godbillon-Vey class and \textsl{characteristic classes of flat bundles} (which include the Chern-Simons invariants for flat connections) is treated in  detail in \cite{Kamber}.

\section{Confoliations}

We shall mention some facts concerning \emph{contact structures}. An orientable closed smooth $(2n+1)$-manifold $M$ is \emph{contact} if and only if there exists a 1-form $\Omega $ such that
$$\Omega \wedge (d\Omega )^{n}\neq 0$$
everywhere on $M$. If we choose local coordinates $(x^{1},...,x^{n},y^{1},...,y^{n},z)$ on $M$, then locally one has
$$\Omega =dz-\sum _{i=1}^{n}y^{i}dx^{i}$$
The $(2n+1)$-manifold $M$ is called \emph{almost contact} if the tangent bundle $TM$ can be reduced to a $U(n)$-bundle over $M$. Contact implies almost contact. There is a topological obstruction for the existence of an almost contact structure which involves the \emph{3rd Stiefel-Whitney} class: it has to vanish in order that the tangent bundle can be reduced to a $U(n)$-bundle (hence the 3rd Stiefel-Whitney class is the primary obstruction for the existence of a contact structure). Using the contact 1-form $\Omega $ one can define a \emph{Poisson} bracket for vector fields on $M$, just like for the symplectic case (see \cite{Gr}).\\

Let now $M$ be a closed smooth oriented \emph{3-manifold}. A codim-1 foliation on $M$ is defined locally by a nowhere vanishing 1-form $\Omega $. Integrability means that 
$$\Omega \wedge d\Omega =0$$
If \textsl{on the contrary} the above expression is everywhere \emph{nonvanishing} on $M$, then we say that $\Omega $ defines a \emph{contact structure} on $M$. A contact structure is the odd-dimensional analogue of a symplectic structure and from its definition we see that it is \textsl{exactly the opposite} of a foliation.\\

Following \cite{Th}, a \emph{positive} (resp. \emph{negative}) \emph{confoliation} is defined by a nowhere vanishing 1-form $\Omega $ on $M$ such that 
$$\Omega \wedge d\Omega \geq (resp. \leq )0$$
This can be generalised for codim-1 foliations on any \emph{odd} dimensional manifold (see \cite{Th}). Then one has the following definition:\\

Let $\Omega $ be a 1-form on $M$ such that $\{\Omega =0\}$ defines a codim-1 foliation $F$ on $M$ (namely $F$ is a codim-1 integrable subbundle of the tangent bundle $TM$ of $M$). We say that $F$ can be \emph{linearly deformed} into a positive contact structure if there exists a deformation $F_t$ given locally by $\{\Omega _{t}=0\}$, where $t$ is real and non-negative such that $\Omega _{0}=\Omega $ and
$$\frac{d(d\Omega _{t}\wedge \Omega _{t})}{dt}|_{t=0}>0$$
The above inequality is equivalent to the inequality
$$<\Omega ,X>:=\Omega \wedge dX+X\wedge d\Omega >0$$
where 
$$X=\frac{d\Omega _{t}}{dt}|_{t=0}$$
Note that this condition depends on the foliation $F$ only and not on the choice of the defining 1-form $\Omega $ (recall that $\Omega $ and $f\Omega $ where $f$ is an arbitrary function define the same foliation $F$).\\

If $\Omega $ is a 1-form that defines a foliation, then we want to find a 1-form $X$ such that $<\Omega ,X>>0$. Let us define a real valued symmetric form $<<\Omega ,X>>$ by integrating the 3-form $<\Omega ,X>$, namely
$$<<\Omega ,X>>=\int _{M^3}<\Omega ,X>$$
Stokes' theorem shows that
$$<<\Omega ,X>>=-2\int _{M^3}\Omega \wedge dX=2\int _{M^3}X\wedge d\Omega $$
If $\Omega $ is a \emph{closed} 1-form, this guarantees that $\Omega \wedge d\Omega =0$, hence integrability holds, but in this case $<<\Omega ,X>>=0$ for \textsl{any 1-form} $X$. So a foliation defined by a closed 1-form \textsl{cannot} be \emph{linearly} deformed into a contact structure. Foliations defined by closed 1-forms are homotopic to foliations with no holonomy (for the proof see \cite{Th}).\\

Conversely if there exists a 1-form $X$ which satisfies the above inequality then the deformation
$$\Omega _{t}=\Omega +tX$$
is the required linear deformation which defines contact structures $F_{t}$ given by the equation $\{\Omega _{t}=0\}$ for small $t\neq 0$. We say that a foliation $F$ can be deformed into a contact structure if there exists a deformation $F_t$ beginning at $F_{0}=F$ such that $F_{t}$ is contact for $t>0$. We shall consider \emph{approximations} of a foliation $F$ by contact structures when it will not be clear that this could be done via a deformation.\\

 There are also \emph{non-linear} deformations of foliations to contact structures, we shall give an example: let $F_0$ be the foliation of the 3-torus $T^3$ by the 2-tori $T^{2}\times p$, where $p\in S^1$. If $x,y,z\in [0,2\pi )$ are coordinates on $T^3$, then $F_0$ is given by the equation $dz=0$. Hence for any integer $n>0$ and any real $t>0$ the form
$$a^{t}_{n}=dz+t(cosnz dx + sinnz dy)$$
defines a contact structure on $T^3$.\\

Thurston \emph{conjectures} in \cite{Th} that it is likely that any foliation on an orientable 3-manifold can be deformed (or approximated) into a contact structure (though not necessarily linearly). He has proved a statement which is only slightly weaker.\\

Anyway, the relation all this has with our case is the following: recall that in order to linearly deform a codim-1 foliation given by a 1-form $\Omega $ into a contact structure, one has to find another 1-form $X$ satisfying a certain inequality described above. Let us consider the case where the 1-form $X$ \textsl{itself also defines another codim-1 foliation}. But $\Omega $ and $X$ have to satisfy the inequality
$$<\Omega ,X>=\Omega \wedge dX+X\wedge d\Omega >0$$
Then the foliations defined by $\Omega $ and $X$ are \emph{transverse} and for all $t\in (0,\pi )$ different from $\pi /2$, the 1-form $\Omega cost+Xsint$ defines a contact structure. We call this a \emph{transverse instanton}, namely an interpolation between \textsl{transverse codim-1 foliations}. There is an alternative description using the notion of \emph{conformally Anosov flows}, namely instead of using the contact structure defined by the 1-form $\Omega cost+Xsint$ one can use a vector field $Y$ say, whose flow is conformally Anosov (the definition is the same using transversality of codim-1 plane fields which provide a continuous splitting of the tangent bundle of our 3-manifold). The 3-torus has many conformally Anosov flows.\\

After this nice geometric interpretation of \textsl{some} noncommutative instantons (namely those interpolating between ``transverse'' codim-1 foliations) we would like to finish this section with another remark: contact structures are the odd-dim analogue of symplectic structures, hence it is natural to try to see how they can be related in the case where for example one has a symplectic 4-manifold with boundary and the 3-manifold on the boundary has a contact structure; foliations in most cases, as we have seen, can be approximated by contact structures. It turns out that a close relation indeed exists but this happens only for a special class of contact structures called \emph{tight}. It is very interesting that this special class of tight contact structures is the approximation of a special class of foliations, called \emph{taut}.\\

A contact structure (namely a codim-1 nowhere integrable subbundle $F$ of the tangent bundle) on a 3-manifold $M$ is called \emph{overtwisted} if there exists an embedded disk $D \subset M$ such that $\partial D$ is tangent to $F$ but $D$ itself is transversal to $F$ along $\partial D$. A contact structure is called \emph{tight} if it is not overtwisted. Overtwisted contact structures are very flexible whereas tight contact structures have a lot of rigidity properties.\\

Moving to the other end of the confoliation scale, namely foliations, we call a foliation \emph{taut} if it is different from the (trivial) foliation $S^{2}\times S^{1}$ and satisfies any of the following equivalent properties which are due to Novikov (see \cite{Nov} or \cite{CC}):\\
1. Each leaf is intersected by a transversal closed curve. ({\bf Remark:} \emph{This can be used as a generalisation of the 1-extra dimension in M-Theory compared to strings}, namely instead of just $M^{11}=X^{10}\times S^{1}$ for some 10-manifold $X^{10}$ on which strings live, we require that $M^{11}$ admits a taut codim-1 foliation).\\
2. There exists a vector field $X$ on $M$ which is transversal to $F$ ($F$ is now an integrable codim-1 subbundle of the tangent bundle) and preserves a volume form on $M$\\
3. $M$ admits a Riemannian metric for which all leaves are minimal surfaces.\\

It seems that in some sense tight contact structures correspond to taut foliations, in the sense that they share many common features as we shall see.\\

Let $(M,F)$ be a confoliation and $\omega $ a closed 2-form on $M$. We say that \emph{$\omega $ dominates $F$} if $\omega |_{F}$ does not vanish.\\

For a foliation the existence of a dominating 2-form $\omega $ is equivalent to property 2. above of being taut. For if the vector field $X$ is transverse to $F$ and preserves a volume form say $Y$, then the closed 2-form $\omega :=X\lrcorner Y$ dominates $F$. Conversely, suppose $\omega $ dominates $F$ defined by $\Omega =0$. Then the vector field $X$ such that $X\lrcorner \omega =0$ and $\Omega (X)=1$ is transverse to $F$ and preserves the volume form $Y:=\omega \wedge \Omega $.\\

Suppose now that a 3-manifold $M$ with a positive confoliation $F$ bounds a compact symplectic 4-manifold $(W,\omega )$. We call $(W,\omega )$ a \emph{symplectic filling} of the confoliated manifold $(M,F)$ if $\omega |_{M}$ dominates $F$ and $M$ is oriented as the boundary of the canonically oriented symplectic manifold $(W,\omega )$. If $F$ is a foliation then the orientation condition is irrelevant. Equivalently in the above situation $M$ is called \emph{symplectically fillable}. (Restriction to connected component gives same definitions with the word ``semi-fillable'' instead of fillable)\\

{\bf Proposition 2:}

Taut foliations are symplectically (semi-)fillable.\\
{\bf Proof:}

Let $\omega $ be a dominating 2-form for the taut foliation $F$ defined by $\Omega =0$ on $M$. Set $W:=M\times [0,1]$ and define a closed 2-form $\tilde{\omega }=p^{*}\omega +\epsilon d(t\Omega )$ where $p$ is the projection $p:W\rightarrow M$. When $\epsilon >0$ is small then the form $\tilde{\omega }$ is non-degenerate and dominates $F$ on $\partial W=M\times 0\cup M\times 1$.\\

It is considerably harder to prove that symplectically (semi-)fillable contact structures are tight (see \cite{Th}). Hence the property of symplectic fillability is shared by taut foliations and tight contact structures. We then have the following\\ 

{\bf Crucial fact:}

\emph{Another property which is common in \textsl{tight} contact structures and \textsl{taut} foliations is that only \textsl{finitely} many homotopy classes of codim-1 subbundles of the tangent bundle $TM$ are representable by taut foliations; but also for a closed orientable 3-manifold $M$ only finitely many cohomology classes from $H^{2}(M)$ can be represented as Euler classes of tight contact structures.}\\

The second statement about tight contact structures follows from Thurston's basic inequality (see \cite{Th1})
$$|e(F)[N]|\leq -\chi (N)$$
where $e(F)$ is the Euler class of the codim-1 nowhere integrable subbundle $F$ of $TM$ which is a contact structure and $N$ is a closed embedded orientable 2-manifold $N\subset M$ and $\chi (N)$ is its Euler characteristic.\\

The first statement concerning taut foliations follows from a result of Kronheimer and Mrowka using Seiberg-Witten theory for symplectic 4-manifolds (namely that symplectically semifillable contact structures may represent only a finite number of homotopy classes of plane fields, see \cite{Kr}).\\ 

 The \emph{restriction} for the above result to hold is that \textsl{the 3-manifold $M$ must have $H_{2}(M)/h(\pi _{2}(M))\neq 0$}, where $h:\pi _{2}(M)\rightarrow H_{2}(M)$ is the Hurewicz homomorphism. Only if this restriction holds the 3-manifold admits a tight (moreover symplectically semi-fillable) contact structure.\\

We end this section with the following remarks: symplectic (semi-)fillability is invariant under surgeries of index 1 and 2. and the 3-torus $T^3$ admits infinitely many non-diffeomorphic but homotopic contact structures.\\

Hence to summarise, the point of this section was twofold:

 First we tried to indicate a possible way to normalise our path integral of the previous section (and hence 3-dim Abelian Chern-Simons theory, at least in some special cases) by using the idea that symplectic 4-manifolds (which are symplectic fillings of the boundary 3-manifolds carrying tight contact structures) give cobordisms of 3-manifolds with codim-1 taut foliations (where codim-1 taut foliations are approximated by tight contact structures). As we saw, interestingly enough, when the topology of the 3-manifold is such that it has a 4-manifold as its symplectic filling, (see the restriction mentioned above), \emph{there are only a finite number of these}. That assures convergence of the series giving the contribution to the path integral from interpolation between taut codim-1 foliations of our 3-torus and hence \textsl{can be used to define quantum invariants for 3-manifolds}. In the next section we shall try to apply some of these ideas to M-Theory in D=11 but in higher dimensions nothing is known; actually even the fact that the simplest 11-manifold, ie $S^{11}$ admits 5-plane fields was only conjectured in \cite{Zois}, the motivation coming from S-duality in M-Theory.

Secondly we tried to exhibit some nice geometric interpretation of noncommutative instantons between transverse codim-1 foliations which can be described by conformally Anosov flows and to underline some striking similarities between taut foliations and tight contact structures; perhaps then these two can be united with a notion of tight (or taut) confoliations and hence one will then have \emph{tight} (or taut) noncommutative instantons (ie interpolation between taut foliations). The right definition is still unsettled in the literature. Moreover taut foliations play an important role if one wishes to study \textsl{metric} aspects of foliations.

\subsection{A new invariant for 3-manifolds?}

It is we believe quite obvious now from our discussion above that we have an \emph{invariant for 3-manifolds} which is just the sum of the Godbillon-Vey classes (integrated over the fundamental class of our 3-manifold) of all the codim-1 \emph{taut} foliations of our 3-manifold. The Godbillon-Vey class for codim-1 foliations is a 3-dim cohomology class hence integrated over the fundamental class of our 3-manifold gives a number. The crucial fact are the Thurston and Kronheimer-Mrowka results that there exist a \emph{finite number} of taut codim-1 foliations (provided of course that the 3-manifold satisfies the requirement stated above for the existence of symplectic fillings), combined with Thurston's theory of confoliations and approximations of taut codim-1 foliations by tight contact structures. \textsl{Hence the sum of all these is finite}. For the moment we are rather careful and we do not claim that this is definitely a \emph{new} invariant for 3-manifolds. (That justifies the question mark in the title of this subsection).\\

A physical interpretation of this is that it corresponds to the condribution to the path integral from interpolation between ``taut'' noncommutative vacua according to the Connes-Douglas-Schwarz interpretation (which are represented by taut codim-1 foliations of the 3-torus or any 3-manifold in fact).\\

The interesting point would be to see its relation to the Chern-Simons topological quantum field theory invariants and everything that follows from them; we suspect that at least with the Abelian Chern-Simons theory invariants must be a close relation. The notion of symplectic fillability also strongly suggests perhaps a relation with symplectic 4-manifolds invariants. Let us underline that this invariant depends \emph{only on the topology of the 3-manifold}, the taut codim-1 foliations are used as an intermediate step which at the end are ``integrated out'', as physicists would say (it's like the use of bundles in Donaldson invariants' case or Jones-Witten invariants' case). The crucial fact is that this ``integration'' actually reduces to a \emph{finite sum}, hence no convergence questions arise, for taut codim-1 foliations, due to  the Thurston and Kronheimer-Mrowka results stated above (although we do not perform any explicit calculations in this article).\\ 

Let us also remark that in the Chern-Simons theory (see \cite{WN}), there is an important restriction: the topology of the 3-manifold is such that there is a finite number of irreducible representations of the fundamental group (hence a finite number of gauge equivalence classes of flat connections). In our case the restriction on the 3-manifolds comes from what is assumed in order to have symplectic fillings (see the last remarks in the previous section). We do not know which  restriction is more narrow. We belive however that perhaps the most interesting point is that this foliation approach makes contact of ``classical topology'' with ``noncommutative topology''; the later seems to provide quite impressive generalisations.

\section{Linearised M-Theory}

Let us start by saying that the reason why we elaborated extensively on the dim 3 case, apart from the definition of a possibly new invariant for 3-manifolds and our new idea to normalise Abelian Chern-Simons topological quantum field theory using \emph{symplectic} 4-manifolds as providing cobordisms of 3-manifolds with \emph{taut codim-1 foliations}, is the hope that improving our understanding in dim 3 may be of some use in treating M-Theory in D=11; in \cite{Zois} a purely topological Lagrangian density was suggested for M-Theory which was in fact the generalisation of the Godbillon-Vey class (which originally applies to codim-1 foliations), that describes codim-5 foliations on an 11-manifold. The motivation for that was the idea that starting from 5-branes (which are the ``solitons'' of D=11 SUGRA) we assumed that our 11-manifold was actually foliated by codim-5 submanifolds, the world volumes of the 5-branes (a 5-brane in time sweps out a 6-manifold, hence its codim is 11-6=5). Our understanding has improved: as we have explained above the 1-form $\theta $ appearing in the Godbillon-Vey class is actually a basic connection on the \emph{normal bundle} of the foliation; its relation to the fundamental 3-form field $C$ of D=11 supergravity was described in \cite{Zois}. But the assumption that the 11-manifold had to be foliated was ad hoc.\\

Now we have a good explanation for that: recall that the Polyakov action for strings actually describes \emph{embeddings} of Riemann surfaces (source space in the $\sigma $ model language) into the target space which in string theory is a 10-manifold. Foliations describe in general \emph{immersed} (6-dim in our case) submanifolds (the leaves, which are the world volumes of the 5-branes) into the target space which now is an 11-manifold (for this elementary but fundamental fact one can see \cite{Hirsch} or \cite{CC}). So by assuming that the target space in M-Theory is foliated by codim-5 submanifolds we essentially immitate the string theory (actually $\sigma $ model) recepie but with one important difference: instead of embeddings we assume immersions! But here comes the crucial fact: in \cite{God} a \emph{calculus for functors} was developed using the notion of the \textsl{differentiation} of a functor; roughly the differential of a functor means the \emph{``best first order linear approximation''}. We shall give a brief outline of the relevant ideas:\\ 

A \emph{prespectrum} is a sequence of based spaces $\{C_{i}|i\geq 0\}$ along with based maps (the \emph{structure maps}) $C_{i}\rightarrow \Omega C_{i+1}$ where $\Omega $ denotes now based loops. A \emph{spectrum} is a prespectrum in which the structure maps are (weak homotopy) equivalences. The \emph{associated spectrum} of a prespectrum $\{C_{i}\}$ is the homotopic colimit $\{hocolim_{j}\Omega ^{j}C_{j+1}\}$.\\

A functor from spaces to based spaces is \emph{linear} if it respects homotopy invariance, it is excisive and also it is reduced; The meaning of the first condition is obvious, the second requirement means that when applying homotopy groups one has a \textsl{Mayer-Vietoris} type of sequence, or in a more advanced terminology a functor is excisive if it takes (homotopy-) co-Cartesian square diagrams to (homotopy-) Cartesian square diagrams; in this case, if we denote our functor $F$, the \textsl{reduced} functor $\tilde{F}(Y)$ applied on a space $Y$ is the fibre of $F(Y)\rightarrow F(*)$, where $*$ is the one-point space. The functor $F$ is called \textsl{reduced} if $F(*)$ is contractible (i.e. can be reduced to a point, hence $F(*)\cong *$). If $F$ is a linear functor, then $\{F(S^i)\}$ is called its \emph{coefficient spectrum}. If $F$ is not excisive, then the spaces $\tilde{F}(S^i)$ still form a prespectrum. The associated spectrum is written $\partial F(*)$ and is called the \emph{derivative} of $F$ at $*$. The \emph{differential of $F$ at $*$}, written $D_{*}F$ is the functor taking a space $Y$ say to the (homotopy co-)limit of
$$\tilde{F}(Y)\rightarrow\Omega \tilde{F}(SY)\rightarrow \Omega ^{2}\tilde{F}(S^{2}Y)\rightarrow ...$$
where $SY$ is the suspension of $Y$. The above discussion can be generalised to the differential of a functor $F$ \textsl{with respect to a space} $X$ instead of $*$, and that is denoted $D_{X}F$. For more details see \cite{God}.\\

Now we apply Goodwillie's general formalism to the \textsl{embedding functor} and we try to approximate it using the \textsl{immersion functor}, this approach is due to M. Weiss (see \cite{Weiss}). So the result is that
the embedding functor (which  assigns to any manifold say $M$ all its embedded submanifolds) is not linear but the immersion functor is linear. But one has more than that: one can think of the \emph{immersion functor as being the linearisation of the embedding functor} or in other words \textsl{one can think of the immersion functor as the \emph{``differential''} of the embedding functor}.  Let us also add that based on Goodwillie's original ideas again, in \cite{Weiss} a \emph{Taylor series} expansion for the functor of embeddings was proposed using immersion theory; moreover the question of \textsl{convergence} was also addressed; \emph{that suggests that one might be able to go further than this linear approximation for M-Theory}.\\ 

Linear functors are particularly useful in topology because \emph{every generalised homology theory (like K-Theory for example) arises from some linear functor of spaces} (this is a classical result due to Whitehead). The above argument is based on a specific application of Goodwillie's work used by Madsen and Tillmann in their programme to prove the \emph{Mumford conjecture} for the (linearised) mapping class group. The justification for this approximation is supported also from the fact that as was pointed out clearly in \cite{Zois} (section 7) in dimensions grater or equal to 4 (namely for $p$-branes where $p\geq 3$) one \emph{cannot} have something analogous to strings (=1-branes) where a summation over all 2-manifolds is performed by summing over all possible gena (for 3-manifolds, namely 2-branes the question from the point of view of topology is still open).\\

Just for convenience we briefly recall that in string theory one starts with harmonic maps $\phi :\Sigma _{g}\rightarrow M^{10}$ which describe embeddings from a Riemann surface $\Sigma _{g}$ with genus $g$ to a target space $M^{10}$ which is a 10-manifold. These 2-dim quantum field theories can be viewed as path integrals $\int D\phi e^{-S[\phi ]}: =F_{g}(M^{10})$ where $S[\phi ]=\int _{\Sigma _{g}}|d\phi |^{2}$. Then one has to sum over all Riemann surfaces, which means a sum for all $g$ from zero (corresponding to the 2-sphere) to infinity. The reason for this is because 2-manifolds have a very simple topological classification: they are classified according to their genus. For 3-manifolds (namely a theory involving 2-branes) the topological classification is not known. \textsl{For dimensions bigger than 3 one \emph{cannot} classify \emph{non simply connected} manifolds.} That means that one cannot go beyond \textsl{tree-level} in perturbation theory. So our approximation involves, for a theory containing say $p$-branes (hence the source spaces will be $(p+1)$-manifolds), summation over all codim-$(n-p-1)$ foliations of the target space, assumed to be an $n$-manifold. It is understood that a single codim-$(n-p-1)$ foliation contains many (in fact infinitely many) $(p+1)$-submanifolds (yet for each one of them one does not have very much control of its topology; they are connected though and their disjoint union gives the target space). Moreover these submanifolds are in general \emph{immersed} and not embedded. Yet this Taylor series expansion may provide a way to go around this problem (the reason that this might work is that a foliation contains many immersed submanifolds, the leaves, and one does not treat each submanifold idividually).\\

So what we actually propose is a \emph{linearised M-Theory} (or perhaps an ``on-shell'' M-Theory). Yet, from the point of view of physics now, this is actually not too bad because no \emph{direct} approach to M-Theory is known; all approaches are via its limiting theories, ie either strings in D=10 or supergravity in D=11. Our approach is addmitedly a linearised version but it is at least direct. Moreover it's a field theoretic approach although our basic field $\theta $ which is a 1-form has a rather complicated relation with the basic 3-form field $C$ of D=11 supergravity (this was exhibited in \cite{Zois} section 7). Yet if one wants to calculate the path integral using only the generalised Godbillon-Vey class as a Lagrangian density, one has to use yet another approximation because the generalised Godbillon-Vey class which describes codim-5 foliations is an 11-form but it is \emph{not quadratic} in the basic field $\theta $ (basic connection on the normal bundle). But things are perhaps much worse because as pointed out to us by A.S. Schwarz the most obvious approximation, namely stationary phase approximation is also useless in this case due to the large power $5$ of the Godbillon-Vey class in this codim-5 case. Nevertheless we think that this should lead to some topological quantum field theory.\\

The Euler-Lagrange equations for the action using the Godbillon-Vey class
$$L=\int _{M^{11}}\theta \wedge (d\theta )^{5}$$
read
$$(d\theta )^{5}=0$$
Particular solutions are \emph{closed} 1-forms $\theta $ and possibly zero is the only minimum of the action.\\

Trying to think of an alternative instead of the Godbillon-Vey class to be used in this 11-dim M-Theory case, another suggestion perhaps would be, if one believes this linear approximation using foliations, the following expression (which is again an 11-form):
$$L=\Omega \wedge d\Omega $$
where $\Omega $ is a decomposable 5-form defining the codim-5 foliations. This Lagrangian density \emph{is quadratic}. The fact that we consider foliations (namely integrable subbundles of the tangent bundle) actually means that our Lagrangian density \textsl{is zero} as we explained earlier. So we should think of foliations as structures defining our theory \emph{on shell}. At this stage we would like to remind one of our last comment in section 4 about the relation between foliations and on-shell Abelian Chern-Simons theory.\\

 At the \textsl{quantum level} though one would expect that quantum corrections will \emph{spoil} this \textsl{foliation structure} and in the relevant path integral one should integrate over \emph{all} 5-forms $\Omega $. The path integral then can be computed directly from our general discussion about degenerate quadratic functionals in section 4 (this functional is obviously degenerate having invariance under addition of exact 5-forms):
$$Z(M^{11})=\int D\Omega exp(i\int _{M^{11}}\Omega \wedge d\Omega)=D(\Delta _5)^{-1/4}D(\Delta _4)^{3/4}$$
using the zeta function regularisation $D$ for the determinant of the corresponding Laplacians. It is perhaps worth mentioning again that now we are not interested in interpolations between homotopically distinct codim-5 foliations, so we integrate over all 5-forms $\Omega $. On shell, these 5-forms have to be decomposable and to satisfy the integrability condition (the action itself then vanishes). We cannot say anything for the normalisation of the path integral. This partition function then is metric independent hence we have a \emph{topological quantum field theory} (the result follows from our discussion in section 4 and essentially it is due to the fact our manifold is \emph{odd} dimensional, in fact 11-dimensional).\\

Note that the Godbillon-Vey class was used as a Lagrangian density to describe contribution from interpolation between non homotopic codim-1 foliations of the 3-torus whereas for this linearised M-Theory (starting from D5-branes) we propose a Lagrangian density involving the \emph{defining equation} for codim-5 foliations. For codim-1 foliations on a 3-manifold, the path integral would be \textsl{the same} no matter if one takes the Godbillon-Vey class or the defining equation as the Lagrangian density assuming integration over \emph{all} 1-forms.\\

The justification for ``switching''---according to what is more convenient for our purposes---from the \textsl{Godbillon-Vey class} (and its generalisations) of the foliation (which involves basic connections on the \emph{normal bundle} of the foliation) to the \textsl{defining equations} of the foliation (involving the actual integrable subbundle of the tangent bundle \emph{itself} which defines the foliation)  comes from the following crucial fact from topology:\\

{\bf Theorem:}\\ 
(We shall not state the theorem in full generality, we shall only mention it in the form which is adequate for our discussion). Fix a smooth closed $n$-manifold $X$ as underlying space. Then there exists a continuous functor
$$\nu:\Gamma _{q}(X)\rightarrow GL(q, {\bf R})$$
defined by
$$\nu (\gamma ^{x})=d\gamma ^{x}$$
the Jacobian at $x$ of any local diffeomorphism whose germ is $\gamma ^{x}\in Hom(x,y)$. This gives rise to the following very important continuous map
$$B\nu : B\Gamma _{q}(X)\rightarrow BGL(q, {\bf R})$$
Hence in any homotopy commutative diagram

\begin{equation}
\begin{CD}
X@>g>>B\Gamma _{q}(X)\\
@V\cong VV     @VVB\nu V\\
X@>>f>BGL(q, {\bf R})\\
\end{CD}
\end{equation}
$g$ is the classifying map of an element of $\Gamma _{q}(X)$ whose normal bundle is classified by $f$.\\

Let us explain the notation: $\Gamma _{q}(X)$ is the topological category of codim-q Haefliger structures on $X$, $B\Gamma _{q}(X)$ is its classifying space and $BGL(q, {\bf R})$ is the classifying space of the topological group $GL(q,{\bf R})$. Over ${\bf R}^{q}$ we construct the sheaf $Sh$ of germs of local diffeomorphisms of  ${\bf R}^{q}\rightarrow {\bf R}^{q}$. That is if $x \in {\bf R}^{q}$, the stalk $Sh_{x}$ is the set of germs at $x$ of diffeomorphisms of open neighborhoods of $x$ onto open sets of  ${\bf R}^{q}$. If $x,y \in {\bf R}^{q}$ we denote $Hom(x,y)$ the set $\{\gamma \in Sh_{x}:\gamma (x)=y\}$.

The proof of this theorem can be found in \cite{Bott} but in fact it is a consequence of \cite{Segal}. See also \cite{Haefliger}. More explanation on the notation can be found in the Appendix (last section).\\

Returning to physics now, we have a nice geometric interpretation: classical (linearised) M-Theory (on shell) corresponds to codim-5 foliations and quantum corrections correspond to deformations! One is tempted to interpret quantum fluctuations loosely as ``confoliations''. Strictly speaking this is not true since confoliations have been defined by Thurston only for codim-1 foliations on 3-manifolds. His definition can be directly generalised to codim-1 foliations on any $n$-manifold but for higher than codim-1 foliations we cannot see immediately how this can be suitably achieved.

Yet now the question that comes forth immediately is the following: what about the Godbillon-Vey class then? It does not play any role at all?

We think it should and in fact there is we believe a third, possibly more interesting choice for a Lagrangian density (still quadratic). But before moving to that, it is quite instructive to have a more detailed look at the non-Abelian Chern-Simons theory from \cite{WN} section 2: ``Weak coupling limit''. 

For convenience we repeat the basic steps, keeping the same notation: starting from the Chern-Simons action on a 3-manifold $M$ say 
$$L=\frac{k}{4\pi }\int _{M}A\wedge dA +\frac{2}{3}A\wedge A\wedge A$$
the stationary phase approximation was followed: the Euler-Lagrange equations state that the connection $A$ is \emph{flat} and then one perturbes a connection around a stationary point, namely $A\rightarrow A_{0}+B$ where $A_{0}$ is flat and $B$ is a small perturbation. Substituting this to the action one gets (keeping terms at most quadratic in $B$):
$$L=kL(A_{0})+\frac{k}{4\pi }\int _{M}B\wedge DB$$
where 
$$L(A_{0})=\frac{1}{4\pi }\int _{M}A_{0}\wedge dA_{0} +\frac{2}{3}A_{0}\wedge A_{0}\wedge A_{0}$$ 
namely the Chern-Simons invariant of the \emph{flat} connection $A_{0}$ and $D$ denotes covariant derivative with respect to the flat connection $A_{0}$. Now this Lagrangian is quadratic in $B$ (ignoring the invariant term for the moment) and one calculates its partition function integrating over \emph{all} 1-forms $B$. The result is the product of determinants of corresponding Laplacians in our terminology of section 4 where we follow Schwarz \cite{S1} (or the ratio of two determinants if one follows the terminology of \cite{WN}) but there is also a contribution (a phase) coming from the Chern-Simons invariant $L(A_{0})$ of the \emph{flat} connection $A_{0}$. Then one has to sum over all flat connections (that's where the crucial assumption mentioned in subsection 5.1 that the topology of the 3-manifold is such that there is only a finite number of gauge equivalence classes of flat connections is used, hence the sum is well-defined).\\

Clearly in our case we would like to think of the \emph{Godbillon-Vey class} as the \textsl{analogue} of the \emph{Chern-Simons invariant} $L(A_{0})$ for the flat connection $A_{0}$ (flat bundles are particular examples of foliations) and our $\Omega \wedge d\Omega $ term is the analogue of the $B\wedge DB$ term integrated over all 5-forms (in accordance to integration over all 1-forms $B$ in \cite{WN}). \textsl{Hence somehow our assumption that there is a foliation structure gives us directly the stationary phase perturbed action.} This seems to be a reasonable choice: on shell, we have a foliation structure and because of this the term $\Omega \wedge d\Omega $ vanishes, but the Godbillon-Vey class does not, so we use that; quantum mechanically, the foliation structure is lost, the Godbillon-Vey class therefore does not exist, but the term $\Omega\wedge d\Omega $ still makes sense and it does not vanish, it's like a deformation of the foliation, or loosely speaking a ``confoliation''.\\

 We may try to immitate \textsl{exactly} the situation in \cite{WN} just described and \emph{twist} our $\Omega \wedge d\Omega $ term by the \textsl{basic} 1-form $\theta $ and get a term which now is (we also assume taking the trace now because forms take values on a vector bundle)
$$L=\int _{M^{11}}\Omega \wedge d_{\theta }\Omega $$
where $d_{\theta }$ is the exterior covariant derivative with respect to the connection $\theta $. Hence on-shell for the 3-dim non-Abelian Chern-Simons theory means a flat bundle, on-shell in our approach means in general a codim-5 foliation.\\

\textsl{We do not know unfortunately how to derive this action in our case from an initial action, namely the analogue of the original 3-dim Chern-Simons action}. Clearly the 11-dim Chern-Simons action does not work because it has high powers in 1-forms (connections) and 2-forms (corresponding curvatures). One may say that in this approach somehow the decomposable (on-shell) 5-form $\Omega $ ``captures'' or ``contains'' as wedge products these elementary 1-forms in a convenient way so that the resulting 11-form $\Omega \wedge d\Omega $ (or $\Omega \wedge d_{\theta }\Omega $) is quadratic (and hence doable). This has certainly some deep relation with integrable systems.\\

Anyway, again following our general discussion in section 4, one can compute the partition function of the twisted term which is:
$$Z(M^{11})=\int D\Omega exp(i\int _{M^{11}}\Omega \wedge d_{\theta }\Omega)=D(\Delta _{\theta }^{5})^{-1/4}D(\Delta _{\theta }^{4})^{3/4}$$
using the zeta function regularisation $D$ for the determinant of the corresponding Laplacians which involve the operator $d_{\theta }$ now and not just $d$ but now acting on forms with \emph{values on the normal bundle of the foliation}.\\

There are more complications arising now though because the forms take values on a vector bundle, which is nonetheless somehow special: its the normal bundle of an integrable subbundle. The problems are basically two: 1. Is the partition function calculated just above \emph{metric independent}? 2. Do we have convergence if we sum over all codim-5 foliations?\\

In \cite{S1} a straightforward  generalisation of real valued forms was considered to the case of forms with values in a \emph{flat bundle}. In this case the partition function gives a \emph{topological invariant}, namely it is metric independent iff the cohomology groups of the 11-manifold $M^{11}$ with values in the flat bundle are trivial.
 
\textsl{We cannot answer either of the two questions above}. All we can suspect is that the analogous restriction in this case perhaps involves the vanishing of the $d_{F}$-cohomology of the manifold or perhaps the condition is that the normal bundle of the foliation must be integrable too with vanishing of the corresponding cohomology (recall that since the subbundle $F$ of the tangent bundle of $M^{11}$ is integrable, that means that the restriction $d_{F}$ of the deRham differential $d$ which takes derivatives along $F$-directions only is also a differential, hence can be used to define a cohomology; in fact it is analogous to the covariant exterior derivative with respect to a connection which takes the ``horizontal component'' of the derivative of a form; the exterior covariant derivative with respect to a connection 1-form is a differential iff the connection is flat). Of course then one would have to sum over all foliations since foliations are assumed to correspond to stationary points of some unknown action. The ``phase'' factor appearing in \cite{WN} due to the Chern-Simons invariant for the flat connection should also be taken into account in the analysis in our case, coming from the Godbillon-Vey invariant of the foliation.  There must not be a continuous family of (homotopically distinct) codim-5 foliations and moreover the sum must converge (recall that in the previous section we considered 3-manifolds with certain topological restrictions so that one had a finite number, hence convergence came for free, of taut codim-1 foliations). We regret to say that a computation as an example is far beyond reach for the moment for it is not even known if the simplest 11-manifold, namely the 11-sphere $S^{11}$, has any codim-5 plane field (namely a codim-5 subbundle of its tangent bundle; if it has one though it will be integrable, hence it will be a foliation; for more on this see \cite{Zois}).\\

The Chern-Simons theory was solved in \cite{WN} using reduction to 2-manifolds and then applying geometric quantisation techniques. In our case that seems to correspond to going down to dim 10 and the only topological methods known there are the Gromov-Witten invariants (topological $\sigma $ models). It seems puzzling that in low dimensions, as was described in the previous section, the existing relation between foliations (and contact structures) is between symplectic manifolds of higher dimension (namely in our case that would be 12, F-Theory perhaps?)\\ 

 Another interesting point might be to try to see \emph{exactly} the relation between K-Theories no1, 2 and 5 in our list in section 1. We would like to underline another interesting coincidence: it is known that the first quantisation of membranes (namely 2-branes which are related by S-duality with 5-branes in M-Theory) faces the impotant problem of having continuous spectrum (just by immitating the string case where a string is seen as an infinite discrete sum of harmonic oscilators). The \emph{leafwise} elliptic pseudodifferential operators may also have, in striking contrast to the elliptic case, continuous spectrum. Perhaps this is more than a simple coincidence. Of course it would be interesting to see if the index theorems for leafwise elliptic operators due to Connes, Moscovici and Skandalis can have any application in M-Theory (these index theorems generalise the original index theorem due to Atiyah for families of elliptic operators).

\section{Appendix}

For convenience we shall give the basic definitions needed to understand some of the material used in this article and to enlighten our definition of K-Theory no2 in our list in the first section.\\

A \emph{Haefliger cocycle} (or $\Gamma _{q}$-cocycle) on a closed smooth $n$-manifold $X$ consists of the following data (note the analogies with the definition of $G$-bundles using $G$-cocycles where $G$ a compact Lie group):\\

1. An open cover $\{U_a\}_{a\in A}$ of $X$.\\
2. For each $a\in A$ a continuous map $f_{a}:U_{a}\rightarrow {\bf R}^{q}$ the germ of which at $x\in U_a$ will be denoted $f^{x}_{a}$.\\
3. For each $x\in U_{a}\cap U_b$ a germ $\gamma ^{x}_{ab}\in Hom(f_{b}(x),f_{a}(x))$ such that:\\

i)  the assignement $x\rightarrow \gamma ^{x}_{ab}$ defines a continuous map $U_{a}\cap U_{b}\rightarrow Sh$\\
ii) $f^{x}_{a}=\gamma ^{x}_{ab}\circ f^{x}_{b}$\\
iii)$\gamma ^{x}_{ab}\circ \gamma ^{x}_{bd}=\gamma ^{x}_{ad}$.\\

We say that two cocycles $c=\{U_{a},f_{a},\gamma ^{x}_{ab}\}_{a,b\in A}$and $c'=\{U_{d},f_{d},\gamma ^{x}_{de}\}_{d,e\in B}$ are \emph{equivalent} if there exists a cocycle corresponding to the covering $\{U_{h}\}_{h\in C}$ where $C$ is the disjoint union of $A$ and $B$ which restricts to $c$ on $\{U_{a}\}_{a\in A}$ and to $c'$ on  $\{U_{d}\}_{d\in B}$. This is an equivalence relation.\\

One then defines the \emph{normal bundle} $Q$ of a Haefliger structure by the $GL(q,{\bf R})$-cocycle $g_{ab}=d(\gamma ^{x}_{ab})$. Equivalent Haefliger cocycles give equivalent (in the sense defined for bundles) $GL(q,{\bf R})$-cocycles, hence the normal bundle of a Haefliger structure (and hence codim-q foliations which are particular examples of $\Gamma _{q}$-structures) is uniquely determined up to isomorphism. For the special case of a foliation $F$ defined by an integrable subbundle $F$ of the tangent bundle $TX$ of our manifold $X$, its normal bundle $Q$ is simply $Q=TX/F$.\\

The space of all $\Gamma _{q}$-cocycles of $X$ modulo equivalence relation is denoted $H^{1}(X,\Gamma _{q})$ and defines a contravariant functor from spaces to sets. Yet this functor is NOT homotopy invariant. In order to get a homotopy invariant functor we impose a further equivalence relation on  $H^{1}(X,\Gamma _{q})$: we say that $a,a'\in H^{1}(X,\Gamma _{q})$ are \emph{homotopic} and write $a\cong a'$ if and only if there exists $b\in H^{1}(X\times I,\Gamma _{q})$ such that $a=i^{*}_{0}(b)$ and  $a'=i^{*}_{1}(b)$. Here of course $i_{0},i_{1}:X\rightarrow X\times I$ are the usual face maps, $I$ is the unit interval and $*$ denotes \emph{pull-back}.\\

If we impose this additional equivalence relation on  $H^{1}(X,\Gamma _{q})$ we obtain another set (in fact topological category) denoted $\Gamma _{q}(X)$; now $\Gamma _{q}(-)$ is a \textsl{homotopy invariant contravariant functor}. That is the functor which Haefliger proved to be representable and mentioned in section 1, K-Theory no2 in our list. Since this functor is representable, one can construct its \emph{classifying space} $B\Gamma _{q}$, hence for any space $X$ one then has a 1:1 correspondence between the set $\Gamma_{q}(X)$ and the set of homotopy classes of maps $[X, B\Gamma _{q}]$. For every space $X$ then $\Gamma (X)$ is a topological category and using this category we apply the Quillen-Segal construction to get the K-Theory $K_{\Gamma (X)}(X)$ which was K-Theory no2 in our list in the first section.\\  

The discussion in this article can be generalised to the study of \emph{codim-q foliations on (2q+1)-manifolds}. This could be an alternative title.\\

\end{document}